\documentclass[aps,prb,twocolumn,superscriptaddress,showpacs,floatfix]{revtex4}

\usepackage{graphicx,color}
\usepackage{dcolumn}
\usepackage{bm}
\usepackage{stmaryrd}
\usepackage{latexsym}
\usepackage{amssymb}
\usepackage{amsfonts}
\usepackage{amsmath}

\begin{document}

\title{Interaction driven quantum phase transition in fractional quantum spin Hall effects}

\author{Wei Li}
\affiliation{State Key Laboratory of Functional Materials for Informatics and Shanghai Center for Superconductivity, Shanghai Institute of Microsystem and Information Technology,
Chinese Academy of Sciences, Shanghai 200050, China}
\affiliation{Department of Physics, State Key Laboratory of Surface Physics and Laboratory of Advanced Materials, Fudan
University, Shanghai 200433, China}

\author{D. N. Sheng}
\affiliation{Department of Physics and Astronomy, California State University, Northridge, California 91330, USA}

\author{C. S. Ting}
\affiliation{Texas Center for Superconductivity and Department of Physics, University of Houston, Houston, Texas 77204, USA}

\author{Yan Chen}
\affiliation{Department of Physics, State Key Laboratory of Surface Physics and Laboratory of Advanced Materials, Fudan
University, Shanghai 200433, China}

\date{\today}


\begin{abstract}
By means of finite size exact diagonalization we theoretically study the electronic many-body effects on the nearly flat-band structure with time-reversal symmetry in a checkerboard lattice model and identify the topological nature of two quantum phases, with ninefold and threefold degeneracy, that appear, respectively, at small and large values $\lambda$ of a nearest neighbor spin dependent interaction. Numerical evidences from the evolution of low-lying energy spectra and Berry phases with both spin-independent and spin-dependent twisted boundary conditions reveal that these two
different ground states share the same topological spin Chern number. Quantum phase transition between these two states by tuning $\lambda$ is confirmed by evaluating the closing of energy and quasispin excitation spectra. At last, the counting rules of spin excitation spectra are demonstrated as the fingerprints of the  fractionalized quantum spin Hall states.

\end{abstract}

\maketitle

{\it Introduction.}\textbf{---}In recent years, the study of $Z_2$ topological insulators with time-reversal invariance has triggered great research activities
both in condensed matter physics and material science~\cite{Hasan,QiZhang,Ando}. Particularly, the two dimensional $Z_2$ topological insulator is a close relative of the integer quantum Hall effect that occurs in semiconductors with sufficiently large spin-orbit coupling and the time-reversal symmetry~\cite{CLKane1,CLKane2}. The prototype model of $Z_2$ topological insulator on honeycomb lattice, Kane-Mele model with $s_z$ conservation~\cite{CLKane1}, can be viewed as a spin dependent version of Haldane lattice model~\cite{Haldane}, namely one could take two copies of Haldane's model with opposite chiralities for up and down spins. This model thus realizes an integer quantum spin Hall effect~\cite{BHZ,Sheng, SCZhang2007}.

Recently, a series of flat-band lattice models with nonzero Chern number, which belong to the same topological class as the Haldane lattice model,
have been proposed~\cite{ETang, KSun, TNeupert}
 and demonstrated to host the fractional Chern insulating phases~\cite{TNeupert,DNSheng,Bernevig,Venderbos1,XLQi,YFWang,DXiao,Zhaoliu,NYYao,WeiLi}
%such a topological band is being fractionally filled by interacting fermions or bosons,
%(such as $\frac{1}{3}$ or $\frac{1}{5}$ filled),
when interacting particles partially fill up these topological flat bands.
Such topological nontrivial states are examples of the fractional quantum Hall (FQH) effect
without  an external magnetic field.
% and may potential persist at very high temperature.
Therefore, there is an intriguing possibility that  a fractional quantum spin Hall (FQSH) effect\cite{Zhang2006, MLevin, Neupert, Neupert2, Repellin}
may also
be realized in the flat-band lattice model as  two copies of the fractional Chern insulators
with opposite chiralities for up and down spin particles, respectively, which may also
survive strong interaction between these spins\cite{strained, kun_yang}.
Interestingly, Neupert {\it et al}~\cite{Neupert,Neupert2} studied the flat-band models  for electron systems
and presented a phase diagram with tunable onsite Hubbard interaction $U$ as well as the nearest neighboring (NN) spin dependent interaction parameter $\lambda$. It is shown that the system favors spontaneously symmetry breaking state when the interaction $U$ dominates,
which leads to the  spin polarized FQH state (Laughlin state~\cite{Laughlin1983}) without
a magnetic field.  On the other hand, when the interaction $U$ and $\lambda$ approach to zero, the system favors two decoupled FQH states, one for each spin orientation, resulting in the ninefold degeneracies at the $\frac{2}{3}$-filling case. With the increase of $\lambda$, the ninefold degeneracies will be lifted and instead threefold degenerated states appear without spontaneously symmetry breaking. However, the nature of the threefold degenerated states and the  phase transition between these
two quantum phases remain not well understood.

In this rapid communication, we systematically study the strong electronic correlation effects on the flat-band checkerboard lattice model with time-reversal invariant by using the finite size exact diagonalization method.
We focus on the nature of two quantum phases, with ninefold and threefold degeneracy, that appear, respectively, at small and large interaction values $\lambda$.
Numerical evidences from the evolution of low-lying energy spectra and Berry phases with both spin-independent and spin-dependent twisted boundary conditions reveal that these two
different ground states share the same topological spin Chern number, thus that both states are FQSH states.  %%structure.
The phase transition can be identified from the closing of energy and quasispin excitation spectra in terms of $\lambda$, besides the change of the topological degeneracy.
Furthermore,  %the fingerprints of both FQSH states,
the counting rules of spin excitation spectra as the evidence of the FQSH states are also studied.

{\it Model and Method.}\textbf{---}We consider the model Hamiltonian of electrons with spin hopping on a checkerboard lattice shown in Fig.~\ref{fig:fig1}(a):

\begin{figure}[tbp]
\includegraphics[bb=10 10 770 380, width=8.7cm,height=4.2cm]{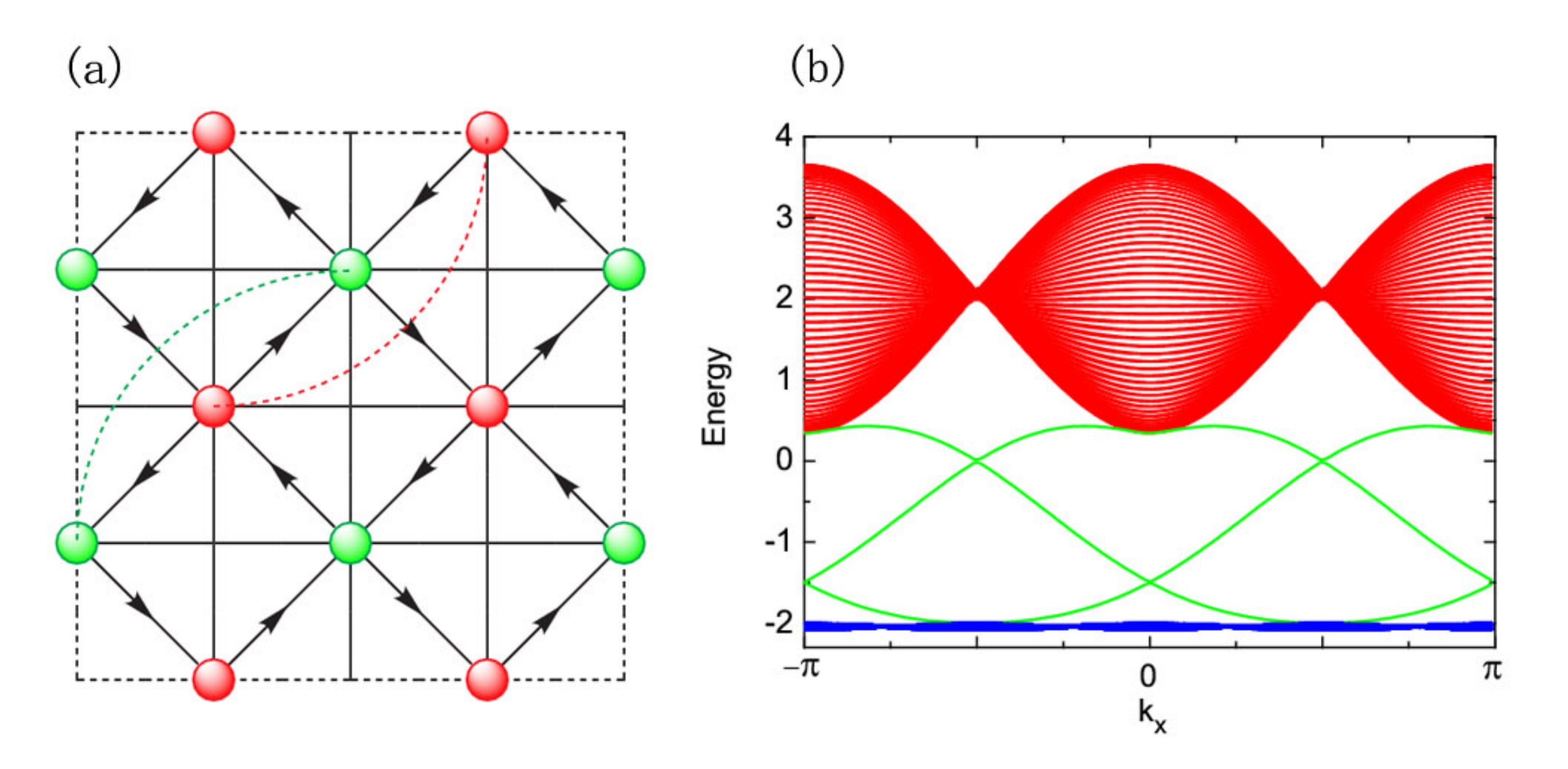}
\caption{(Color online) (a) The checkerboard lattice structure of the flat-band model, with arrows and (solid and dashed) lines representing the NN and NNN hoppings, respectively. The direction of the arrow shows the sign of the phase in the NN hopping terms. Two of the NNN hoppings are shown as the dashed curve. (b) The single-particle energy dispersion of putting the system on a cylinder. The time-reversal invariant chiral edge states (green lines) are observed.}\label{fig:fig1}
\end{figure}

\begin{eqnarray}
{\hat H} &=& {\hat H_{0}} + U\sum_{i}\hat{n}_{i,\uparrow}\hat{n}_{i,\downarrow} + V\sum_{\langle i,j \rangle}[\hat{n}_{i,\uparrow}\hat{n}_{j,\uparrow}+\hat{n}_{i,\downarrow}\hat{n}_{j,\downarrow}
\notag\\
&+& \lambda(\hat{n}_{i,\uparrow}\hat{n}_{j,\downarrow} + \hat{n}_{i,\downarrow}\hat{n}_{j,\uparrow})],
\label{eq:one}
\end{eqnarray}
where ${\hat H_{0}}$ consists of two copies of the $\pi$-flux phase with flat-bands that was proposed in Ref.~\onlinecite{KSun}, one copy for each spin-$\frac{1}{2}$ component preserving  the time-reversal symmetry,
which realizes  the Kane-Mele model for topological insulator. We denote $\hat{c}^{\dag}_{\mathbf{k},\alpha,\sigma}$ as the creation operator for an electron with lattice momentum $\mathbf{k}$ and spin $\sigma=\uparrow,\downarrow$ in the sublattice $\alpha=A, B$ and combine them in the sublattice spinor $\psi^{\dag}_{\mathbf{k},\sigma}=(\hat{c}^{\dag}_{\mathbf{k},A,\sigma},\hat{c}^{\dag}_{\mathbf{k},B,\sigma})$. Then, the second quantized single-particle Hamiltonian reads
\begin{eqnarray}
{\hat H_0} &=& \sum_{\mathbf{k}}\left(
\psi^{\dag}_{\mathbf{k},\uparrow}\frac{\mathbf{B_{\mathbf{k}}\cdot\mathbf{\tau}}}{|\mathbf{B_{\mathbf{k}}}|}\psi_{\mathbf{k},\uparrow} +
\psi^{\dag}_{\mathbf{-k},\downarrow}\frac{\mathbf{B_{\mathbf{-k}}\cdot\mathbf{\tau}^{T}}}{|\mathbf{B_{\mathbf{-k}}}|}\psi_{\mathbf{k},\downarrow}
\right),
\label{eq:two}
\end{eqnarray}
where the three vectors $\mathbf{B_{\mathbf{k}}}$ are defined by
\begin{subequations}\label{Eq.3}
\begin{eqnarray}
B_{0,\mathbf{k}}=4t_3 \cos k_x \cos k_y,
\end{eqnarray}
\begin{eqnarray}
B_{1,\mathbf{k}} + iB_{2,\mathbf{k}} &=& t_1e^{-i\pi/4}(1+e^{i(k_y-k_x)})
\notag\\
&+& t_1e^{i\pi/4}(e^{-ik_x}+e^{ik_y}),
\end{eqnarray}
\begin{eqnarray}
B_{3,\mathbf{k}} = 2t_2(\cos k_x - \cos k_y),
\end{eqnarray}
\end{subequations}
and the identity matrix and the three Pauli matrices $\mathbf{\tau} = (\tau_0, \tau_1, \tau_2, \tau_3)$ act on the sublattice index. Here, $t_1$, $t_2$, and $t_3$ represent the NN, next-nearest-neighboring (NNN), and third-NN hopping amplitudes, respectively. The single-particle band dispersion of the system on a cylinder is shown in Fig.~\ref{fig:fig1}(b). It is clearly shown that there is a large bulk energy gap with the gap amplitude of 2$t_1$ well separating the flat-band and conduction band. It is interesting to point out that there are some edge states emerging within the bulk energy gap and crossing each other at the $\Gamma (k_x=0)$ point forming the Dirac-like dispersion relation protected by time-reversal symmetry. As the bulk energy gap is much larger than the energy scale of the interactions, we can safely project Hamiltonian (\ref{eq:two}) onto the states in the lowest two spin dependent flat-bands
in the exact diagonalization study using the torus geometry. The repulsive interactions in this paper defined in Hamiltonian (\ref{eq:one}) include an onsite Hubbard term $U$ and a NN term which is parameterized by the coupling $V$ and the dimensionless number $\lambda$. It is important to point out that the phase diagram with the effects of $U$ and $\lambda$ with a given finite value $V$ has been constructed in Ref.~\onlinecite{Neupert}. Due to the presence of $U$ term, the system favors spontaneously symmetry breaking state. Thus, we neglect this interaction term and focus on the discussion of the interaction term $\lambda$ effects on the flat-band model with time-reversal invariance.

Next we exactly diagonalize the many-body Hamiltonian as shown in Eq. (\ref{eq:one}) projected to the lowest spin-degenerate flat-bands for a finite system with $N_x \times N_y$ unit cell (total number of sites $N_s = 2 \times N_x \times N_y$) shown in Fig.~\ref{fig:fig1}(a). We denote the number of fermions as $N_e$ ($N_{\uparrow} + N_{\downarrow}$), and filling factor is $\nu = \frac{N_e}{2N_xN_y}$. Because of the periodic boundary condition implementing translational symmetries, we diagonalize the system Hamiltonian in each total momentum $\mathbf{q}=(2\pi k_x/N_x, 2\pi k_y/N_y)$ sector with $(k_x, k_y)$ as integer quantum numbers. Without loss of generality, we set the $t_1$ as an energy unit and the interaction $V=1$. In the following, the filling factor is set to be $\frac{1}{3}$. %, which can be transformed into the filling factor $\frac{2}{3}$ reported in Ref.~\onlinecite{Neupert} by using a particle-hole transformation in the flat-band limit.
Similar results for $\frac{1}{5}$-filling case can also be obtained when the NNN repulsion is included (not presented here).

\begin{figure}[tbp]
\includegraphics[bb=40 10 725 550, width=4.2cm, height=3.2cm]{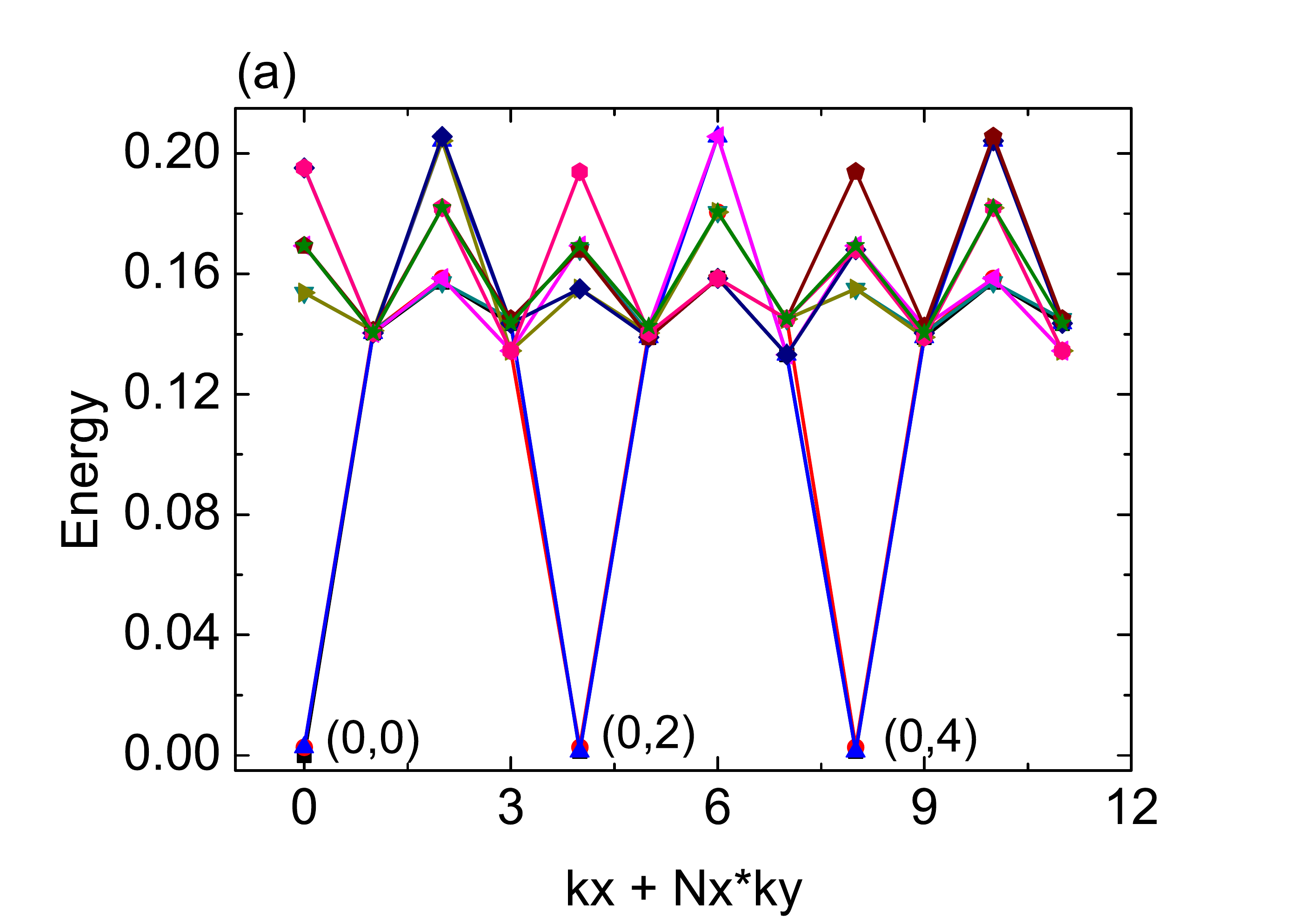}
\includegraphics[bb=40 10 725 550, width=4.2cm, height=3.2cm]{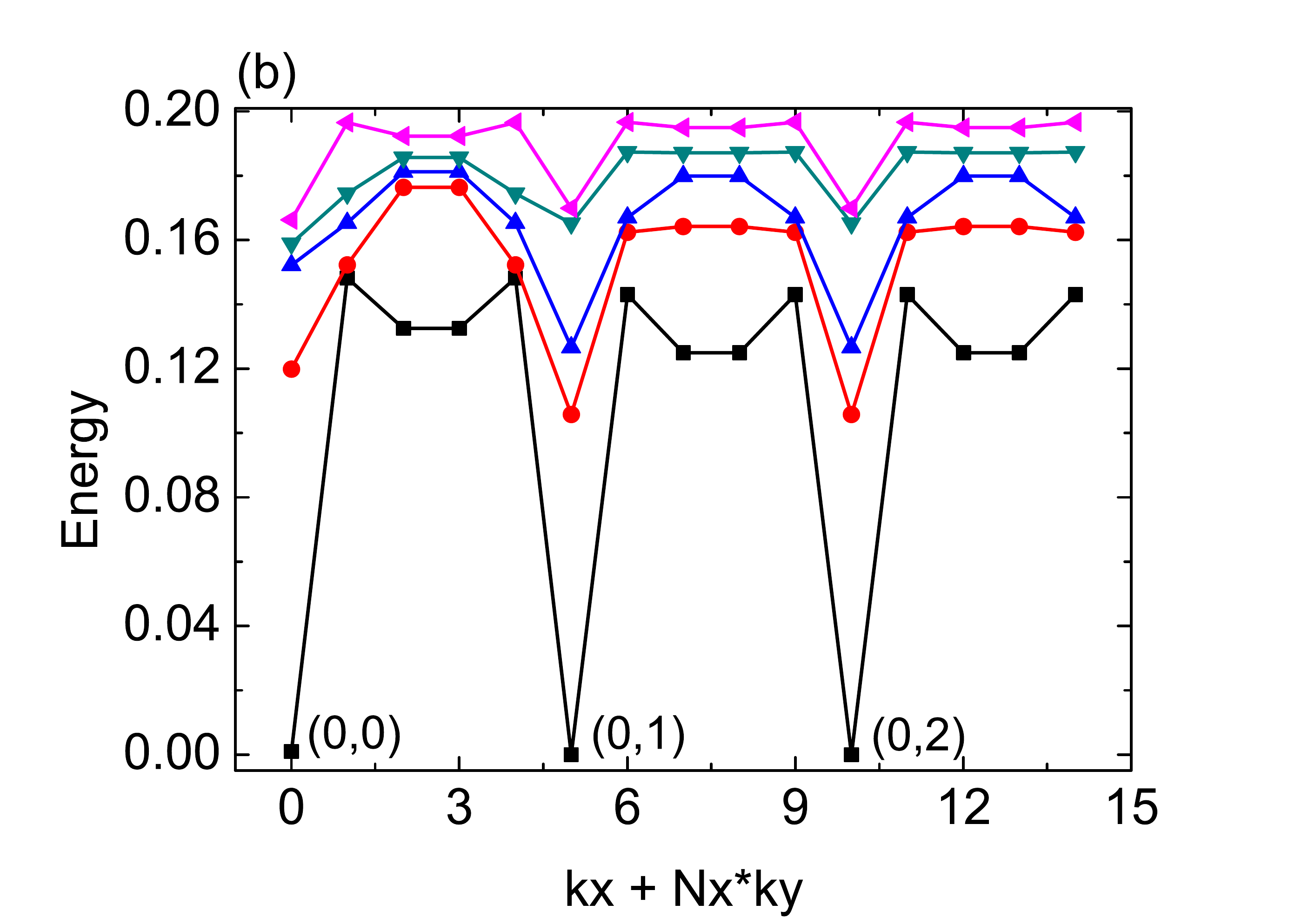}\\
\includegraphics[bb=40 10 725 550, width=4.2cm, height=3.2cm]{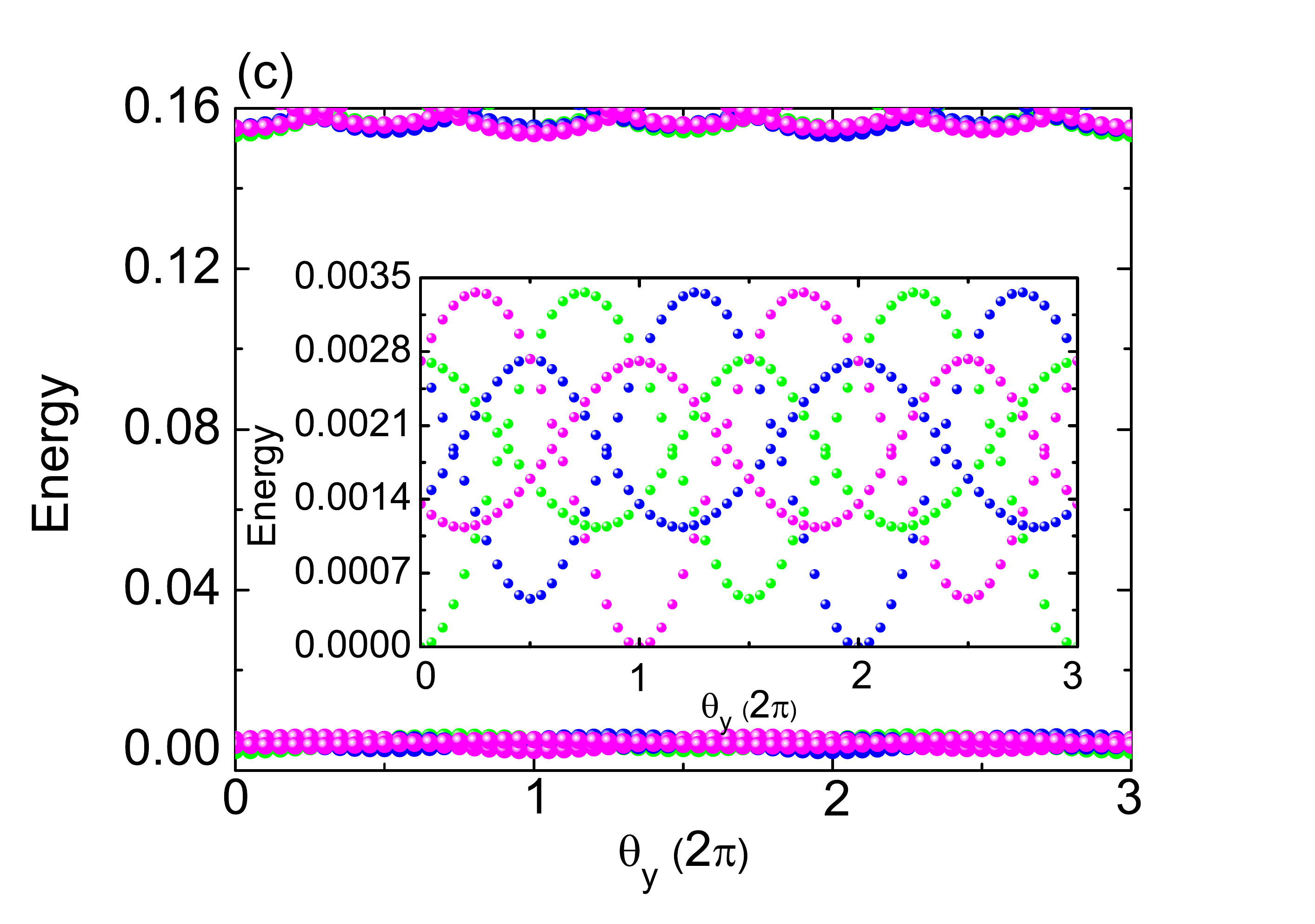}
\includegraphics[bb=40 10 725 550, width=4.2cm, height=3.2cm]{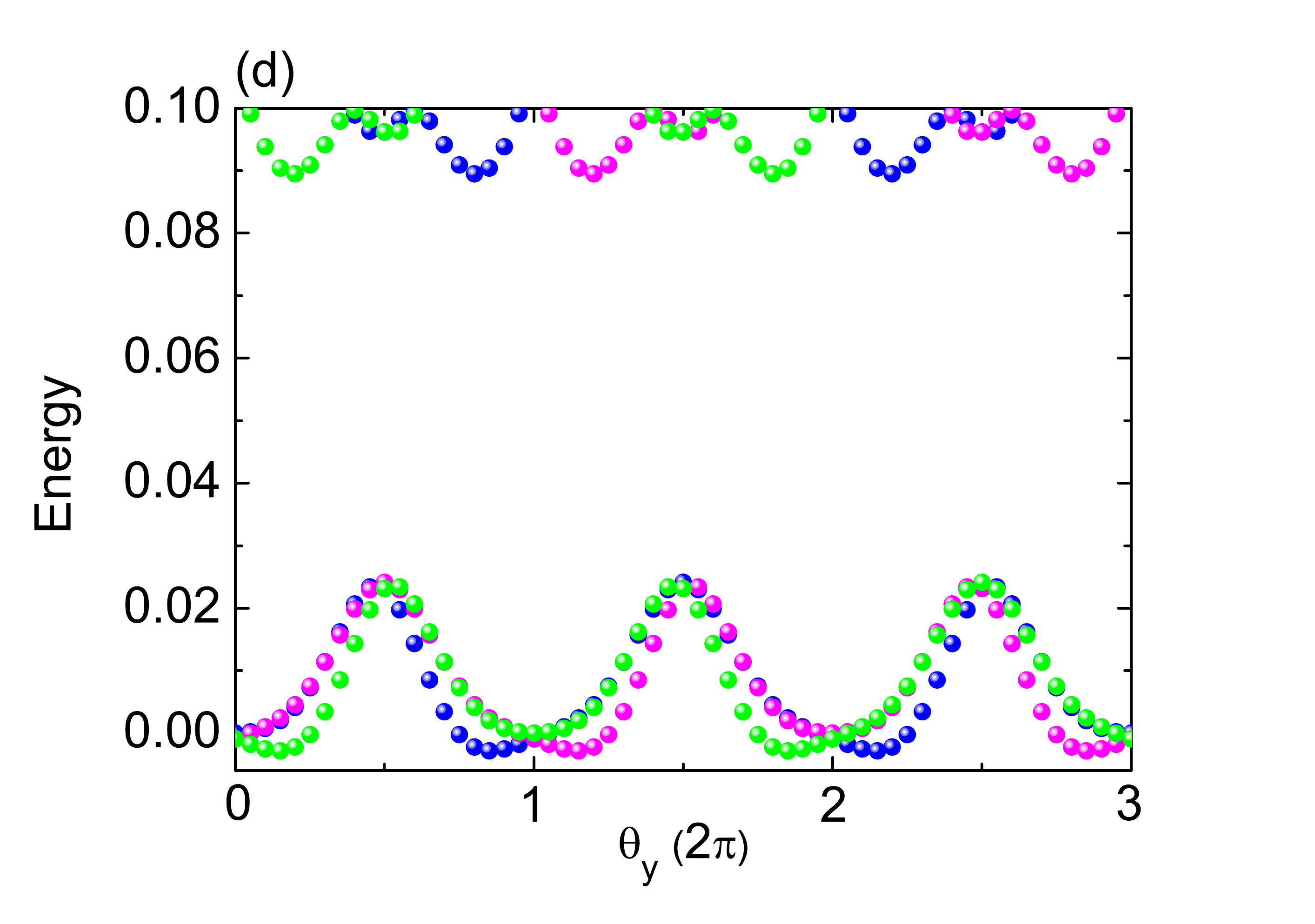}\\
\includegraphics[bb=40 10 725 550, width=4.2cm, height=3.2cm]{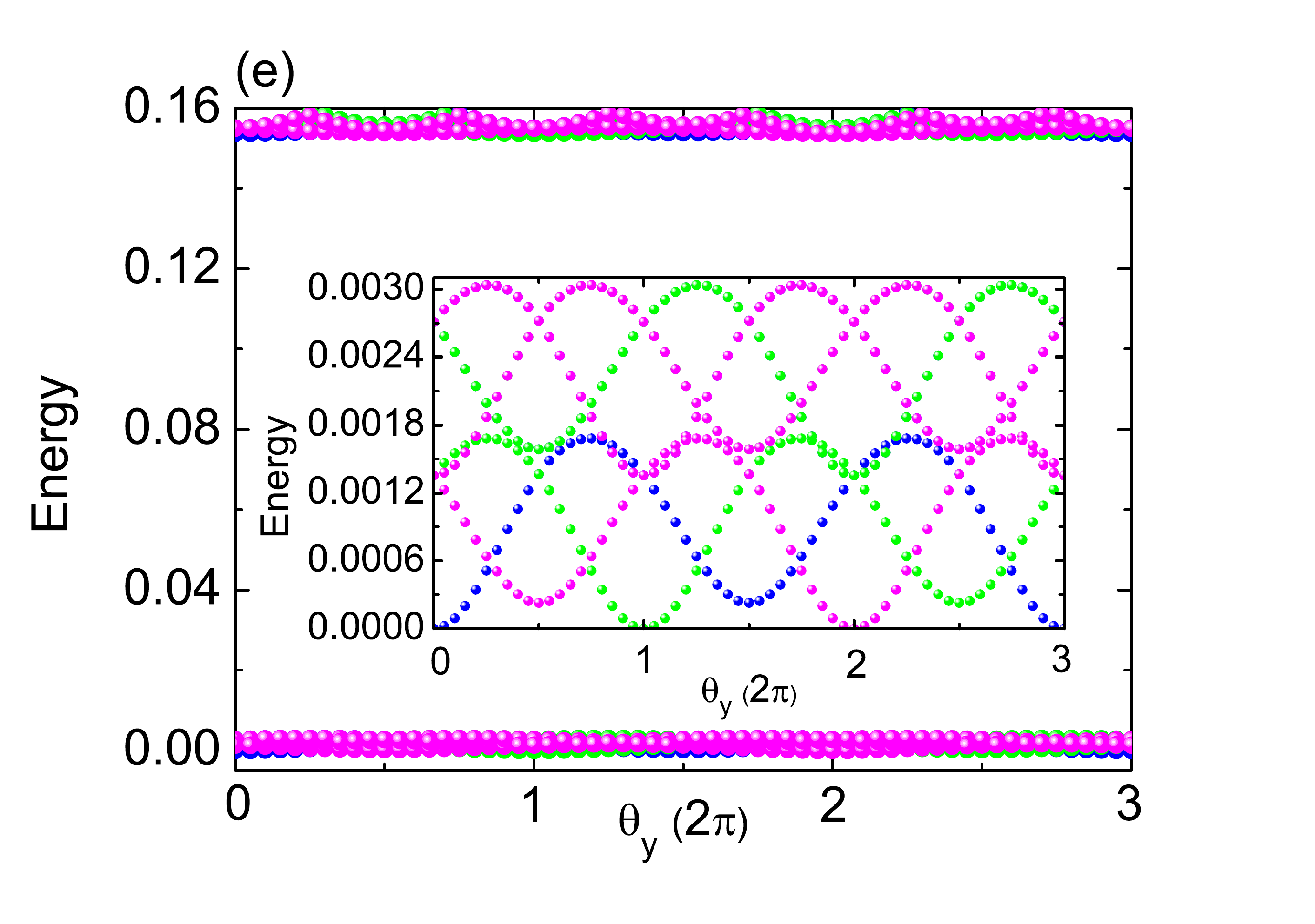}
\includegraphics[bb=40 10 725 550, width=4.2cm, height=3.2cm]{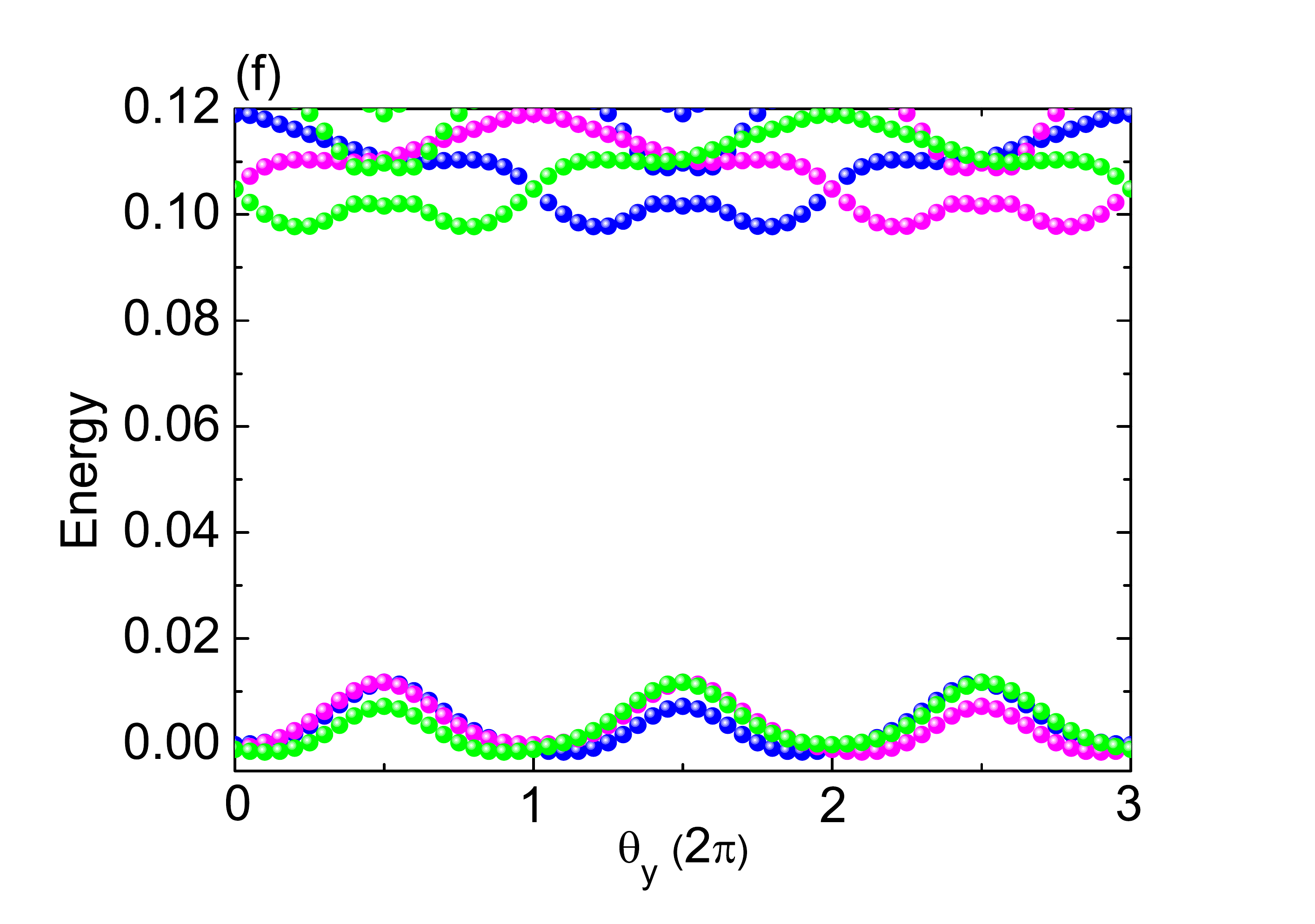}
\caption{(Color online) The interaction $\lambda=0$ and $N_e = 8 (N_s=24)$ for left column while $\lambda=1$ and $N_e = 10 (N_s=30)$ for right column. (a)-(b) Ground state degeneracies.
Evolution of low-lying energy spectra with different twist boundary conditions, (c)-(d) represent spin-dependent twist boundary condition case while (e)-(f) correspond to spin-independent case. Additionally, these ground state energies are all shifted by $E_1$, which is the lowest energy for each system size. }\label{fig:fig2}
\end{figure}

{\it Ground state properties.}\textbf{---}In Figs.~\ref{fig:fig2}(a) and (b), the ground states manifold for the case $\lambda=0$ and $\lambda=1$, respectively, is defined as a set of lowest states [ninefold degeneracies in (a) and threefold degeneracies in (b)] well separated from other excited states by a clear energy gap, which is a necessary condition for the emerging of the fractional Chern insulating states. It is worth pointing out that these two results are the same as that reported in previous theoretical studies for $\frac{2}{3}$ filling~\cite{Neupert}. From Figs.~\ref{fig:fig2}(a) and (b), we notice that the energy gap is always significantly larger than the ground state splitting for various system sizes. Although these states are not exactly degenerate on a finite system, their energy difference should fall off exponentially as the system size increases. In addition, it is interesting to find that for both states with  threefold  or ninefold degeneracy, if $(k_x, k_y)$ is the momentum sector for one of the states in the ground states manifold, then the  other state should be obtained in the sector $(k_x + N_e, k_y + N_e)$ [modulo $(N_x, N_y)$]. This relationship of the quantum numbers of the ground states manifold implies the topological nontrivial characteristics of the Abelian FQH state~\cite{DNSheng, Bernevig}.

To reveal the quantization of the spin trasport for such a topological nontrivial state, we further calculate
the evolution of low-lying energy spectra %with both spin-independent and spin-dependent twist boundary conditions,
by inserting spin independent (dependent) boundary phases coupling to  each spin component  in the system.
For a many-body state~\cite{QNiu,Sheng}: $|\Psi(\mathbf{r}_j)\rangle$, the twisted boundary condition in the $x(y)$ direction is $|\Psi(\mathbf{r}_j+N_{x(y)}\mathbf{a}_{x(y)})\rangle=e^{i\theta_{x(y)}\tau_{3(0)}}|\Psi(\mathbf{r}_j)\rangle$, where $\theta_{x(y)}$ is the boundary phase along $x(y)$ direction and $\mathbf{a}_{x(y)}$ is the lattice vector. According to Laughlin's gauge argument~\cite{Laughlin,Halperin}: for the $\frac{1}{3}$-filling FQH system, when the flux adiabatically inserts three quantum fluxes, the states should evolve back to themselves looking exactly the same as before. From Figs.~\ref{fig:fig2}(c)-(f), we notice that the spin evolution spectrum and the charge evolution spectrum for both threefold and ninefold degenerate states share the same quantization pattern: both the three states and nine states are found to evolve into each other with level crossing and separated from the other low-energy excitation spectrum when imposing the boundary phases. Eventually, all levels return to their initial configuration  after the insertion of three flux quanta, and we also find that each group of three
lowest energy states  shares a  spin Chern number 2 for both phases. This behavior of the spectral flow indicates that the spin Hall conductance is
quantized at $\sigma_H=\frac{2}{3}\frac{e^2}{h}$~\cite{Sheng,TKNN}, which we have obtained by calculating the many-body spin Chern number\cite{Sheng}.

\begin{figure}[tbp]
\includegraphics[bb=50 30 760 560, width=8.5cm, height=7cm]{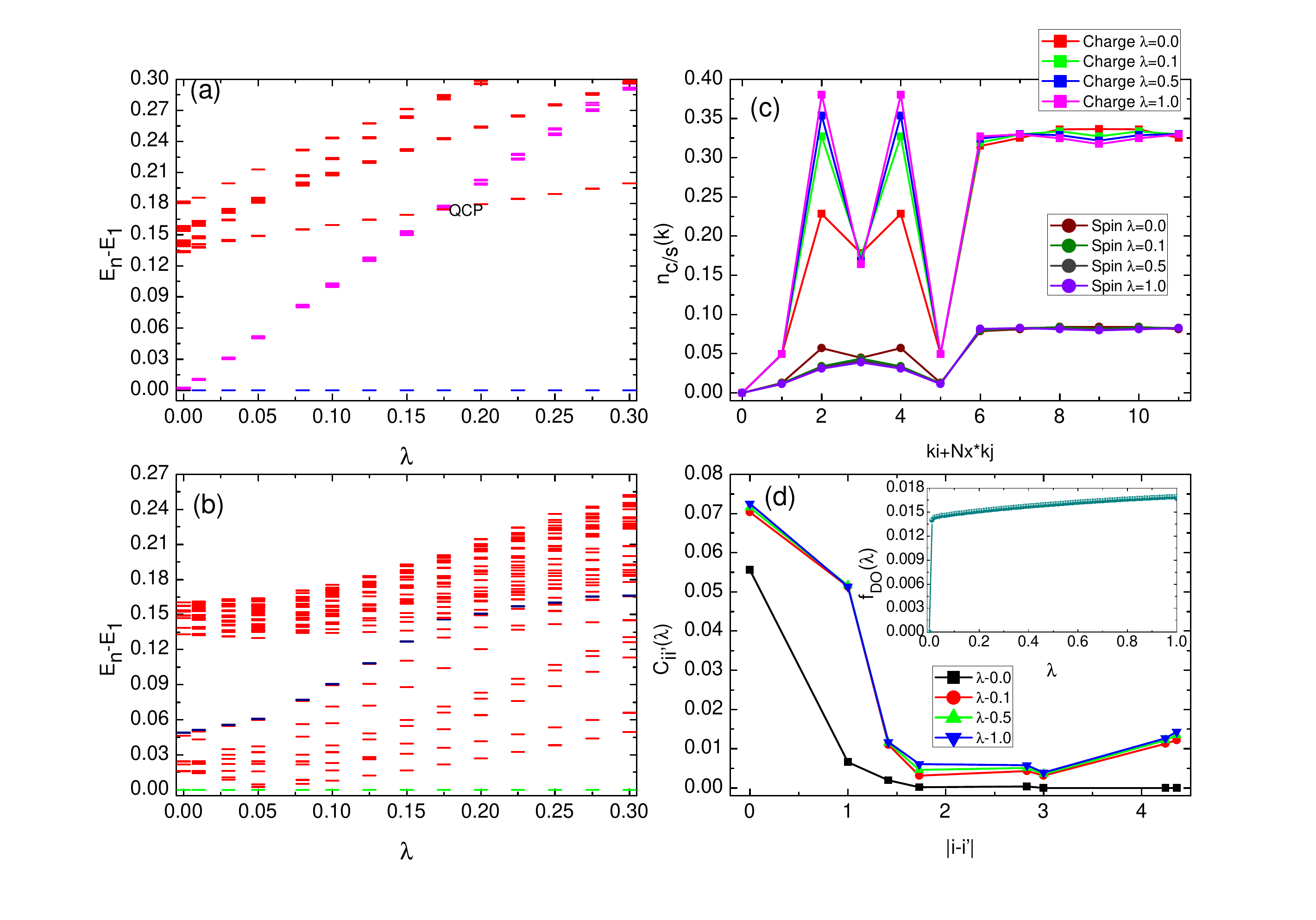}
\caption{(Color online) (a) and (b) describes the sixty lowest eigenenergies [Note: Blue symbols denote the lowest three states, magenta symbols denote the next six states, and red symbols denote the rest excitation states in (a), and green and navy symbols denote the first and tenth state in (b), respectively] and quasispin excitation spectra [at ($k_x$,$k_y$)=(0,0)] of the Hamiltonian in Eq. (\ref{eq:one}) as a function of interaction $\lambda$, respectively. (c) Static structure factor for charge and spin degrees of freedom and (d) pairing correlation function with various interaction $\lambda$. The insert shows the double occupancy $f_{DO}(\lambda)$ as a function of $\lambda$. The system size denotes as $N_s = 2\times N_x(=2)\times N_y(=6)$.}\label{fig:fig3}
\end{figure}

{\it Quantum phase transition.}\textbf{---}Since the two FQSH states have the same spin Chern number
 at $\lambda=0$ and $\lambda=1$, it is unclear what induces the quantum phase transition which changes the ground
state degeneracy of the system.  To address this issue, we calculate the sixty lowest eigenenergies of the Hamiltonian in Eq. (\ref{eq:one}) as a function of interaction $\lambda$ and shown in Fig.~\ref{fig:fig3}(a). We notice that the high ground states degeneracies (ninefold state) will be broken into two sets, one is the lower three states and the other one is higher six states, when the interaction $\lambda$ is turned on. As further increase the interaction $\lambda$, the six states will be raised and crossed into the excitation levels eventually leading to a quantum phase transition. The value of quantum critical point (QCP) in such quantum phase transition denoted in Fig.~\ref{fig:fig3}(a) is about $\lambda\approx0.17$, which is consistent with the value of evaluating the gap closing point of quasispin excitation spectra as a function of interaction $\lambda$ shown in Fig.~\ref{fig:fig3}(b). The detailed counting rules of quasispin excitation spectra will be discussed on the next section.

\begin{figure}[tbp]
\includegraphics[bb=40 50 720 560, width=8.7cm, height=7cm]{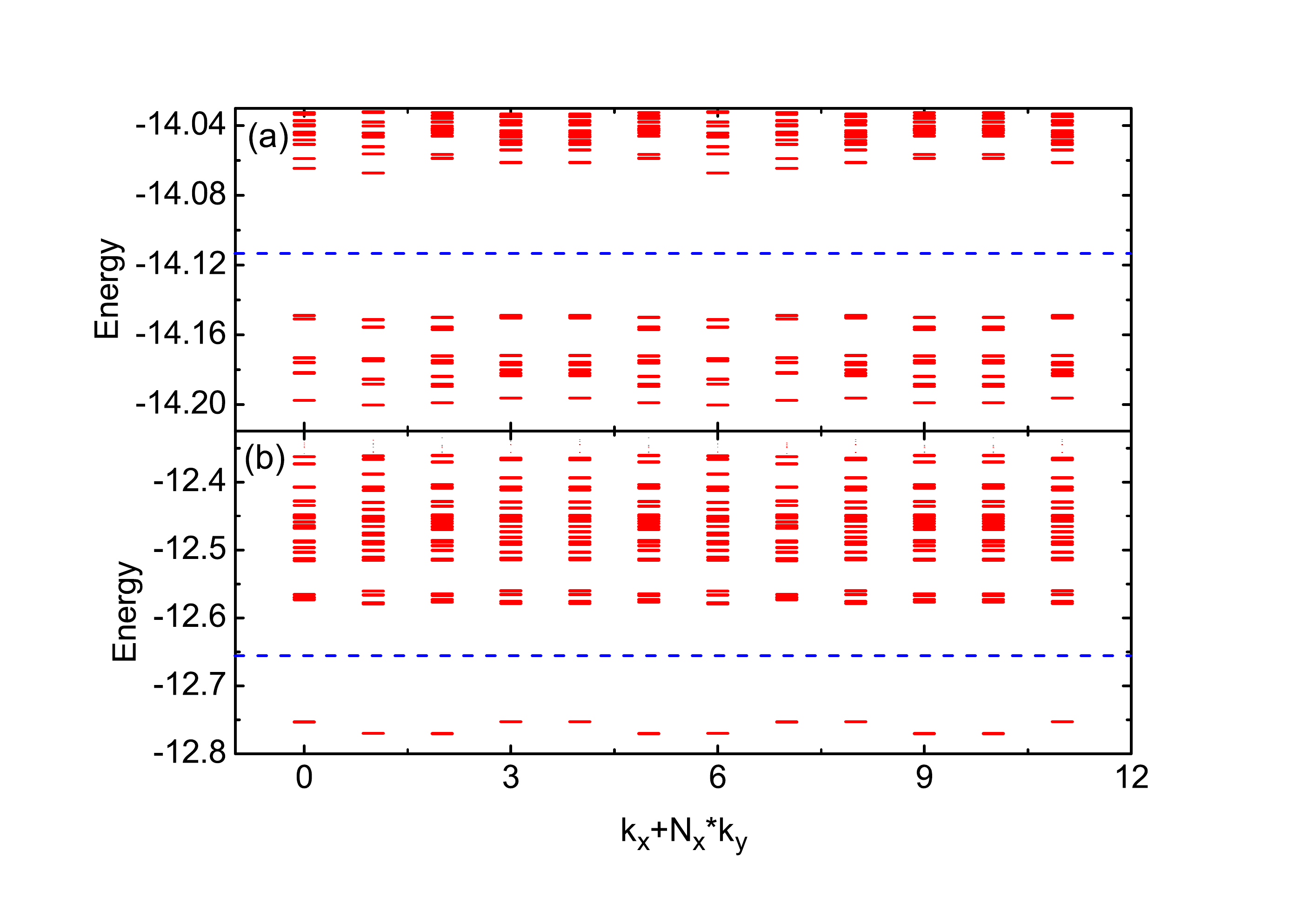}
\caption{(Color online) (a) $\lambda=0$ and (b) $\lambda=1$ for the low-energy quasispin excitation spectrum of the 2D checkerboard model with $N_e = 7$ electrons ($N_{\uparrow}$=3 spin-up and $N_{\downarrow}$=4 spin-down) on the $N_s=24$ lattices. The number of states below the blue dashed line is 10 in each momentum sectors in (a), while the one in (b) is 1 in each momentum sectors.}\label{fig:fig4}
\end{figure}

To reveal the nature of the quantum phase transition and examine whether the system exists a long range order, we study the correlation function of the ground state with various $\lambda$. Here, we first calculate the static structure factor with charge and spin degrees of freedom~\cite{WeiLi}: $n_{c(s)}(\mathbf{k})=\frac{1}{N_s}\sum_{j,l}^{N_s}e^{i\mathbf{k}\cdot(\mathbf{R}_j-\mathbf{R}_l)}(\langle\Psi|\hat{n}(\hat{S})_j\hat{n}(\hat{S})_l|\Psi\rangle-\langle\Psi|\hat{n}(\hat{S})_j|\Psi\rangle\langle\Psi|\hat{n}(\hat{S})_l|\Psi\rangle)$ , where $\hat{S}_i=\frac{1}{2}(\hat{n}_{i\uparrow}-\hat{n}_{i\downarrow})$ is a spin operator, and the wave function $|\Psi\rangle$ is incoherent summation over the degenerate ground states, shown in Fig.~\ref{fig:fig3}(c). It is shown that the pesk values of the correlation functions with charge and spin degrees of freedom are comparable with the average of $\langle \hat{n}_k\rangle$ upon various interactions $\lambda$ indicating the presence of short distance correlations instead of the long-range order % featureless  indicating the absence of long range order
in the system.
Furthermore, we also calculate the pairing correlation function for a finite size system: $C_{i,i'}(\lambda) = \langle\Psi| \hat{\Delta}^{\dag}_{i} \hat{\Delta}_{i'} |\Psi\rangle$, where $\hat{\Delta}_{i} = \hat{c}_{i,\uparrow}\hat{c}_{i,\downarrow}$ is a pairing operator, and find that the pairing correlation function decays rapidly as the distance between any two points is increased [see Fig.~\ref{fig:fig3}(d)]. This   indicates the absence of possible long range ordering for the finite size system of threefold degenerate state ($\lambda = 1$). Thus, we can expect that the system may not exhibit long range charge density wave, spin density wave, and  long-range pairing correlation for large system limit. However, it is interesting to point out that there is a large  value for the same site of pairing correlation function, which is the enhanced  double occupancy. The double occupancy, $f_{DO}(\lambda) = \frac{1}{N_xN_y}\sum_i(\langle\Psi|\hat{n}_{i,\uparrow} \hat{n}_{i,\downarrow} |\Psi\rangle - \langle\Psi| \hat{n}_{i,\uparrow} |\Psi\rangle \langle\Psi| \hat{n}_{i,\downarrow} |\Psi\rangle)$, as a function of $\lambda$ is shown in insert of Fig.~\ref{fig:fig3}(d). The results reveal that the particles with spin-up and spin-down prefer to occupy the same site with the  tuning on
of nonzero $\lambda$. This behavior can be easily understood in the strong coupling limit that the strong NN repulsive interaction $\lambda$ may induce a negative onsite Hubbard interaction.
% $U$ resulting that the particles with different spin orientations favor to occupy the same site.
%In addition, we also checked this point that diagonalize the Hamiltonian in Eq. (\ref{eq:one}) with negative onsite Hubbard $U$ and obtain the same as we expected conclusions.
On the other hand, the induced negative onsite Hubbard interaction is not strong enough to pair the two electrons with different spins locally, the system, thus, does not exhibit long-range ordering. This local spin pairing may also be a consequence of the anti-symmetry of the states for different spins, which intend to form local spin singlet.

{\it Spin excitation spectrum.}\textbf{---}To explore the possible fractional statistics~\cite{Laughlin1983,Halperin1984,Arovas,Leinaas}, we turn to discuss the quasispin excitations, which is one of the most important characteristics for FQSH state. By keeping $N_s$ fixed and changing $N_e$, we can add one quasispin into the system, as shown in Figs.~\ref{fig:fig4}(a) and (b). An energy gap in both figures is clearly visible in the quasispin excitation spectra with the total number of states below the gap differing. In Fig.~\ref{fig:fig4}(a), the number of states below the gap is 120. This result can be easily understood in terms of two decoupled FQH states, one for each spin component obeys the (1, 3)-admissible rule based on the generalized Pauli principle~\cite{Bernevig}. Thus, the total counting gives: $N(\lambda=0) = 3N_xN_y\frac{(N_xN_y - 2N_{\uparrow} - 1)!}{N_{\uparrow}!(N_xN_y - 3N_{\uparrow})!}.$ Putting the system parameters in Fig.~\ref{fig:fig4}(a) into this formula, we get $N(\lambda=0) = 120$, which agrees precisely with the number of states below the spectral gap. In Fig.~\ref{fig:fig4}(b), the number of states below the gap is 12, which is much less than the one in Fig.~\ref{fig:fig4}(a). Because there are $\left(\begin{array}{c}
                   N_{\uparrow} \\
                   1
                 \end{array}\right)=\left(\begin{array}{c}
                   4 \\
                   1
                 \end{array}\right)=4$ configurations emerging for spin up species when removing a spin up particle from the system. Further consider the threefold degeneracies for down spin species and the particles with different spin orientations favor to occupy the same site at the phase ($\lambda=1$) leading to 3 possible extra combinations, we finally get the total 12 quasispin excitation states in accordance with our numerical results.

{\it Conclusion.}\textbf{---}We have performed the exact diagonalization on the study of the electronic many-body effects on the nearly flat-band structure with time-reversal symmetry in a two-dimensional checkerboard lattice model.
The ground state degeneracies of the system can be either ninefold or threefold degeneracies depending on the strength of $\lambda$, both are FQSH states with the same
quantized spin Hall conductance, which can be confirmed systematically by the evolution of low-lying energy spectra with both spin-independent and spin-dependent twisted boundary conditions as well as the many-body spin Chern number calculation. The quantum phase transition from ninefold degenerate state to threefold degenerate one by tuning
$\lambda$ is demonstrated by evaluating the closing of energy and quasispin excitation spectra.
At last, the fingerprints of those FQSH states, the counting rules of spin excitation spectra, are also presented.

{\it Acknowledgement.}\textbf{---}We thank R. B. Tao, Y. S. Wu, X. M. Xie, T. K. Lee, S. Q. Shen, Z. Liu, and Z. C. Gu for helpful discussions. This work was supported by the Strategic Priority Research Program (B) of the Chinese Academy of Sciences (Grant No. XDB04010600) and the National Natural Science Foundation of China (Grant No. 11227902) (W.L.), the State Key Programs of China (Grant Nos. 2012CB921604 and 2009CB929204) and the National Natural Science Foundation of China (Grant Nos. 11074043 and 11274069) (Y.C.), the US Department of Energy, Office of Basic Energy Sciences under grant DE-FG02-06ER46305 (D.N.S.), and the Robert A.Welch Foundation under Grant No. E-1146 (C.S.T.). W.L. also gratefully acknowledges the financial Sponsored by Shanghai Yang-Fan Program (Grant No. 14YF1407100).

\end{document}